\documentclass[journal]{IEEEtran}
\usepackage{graphicx}% Include figure files
\usepackage{dcolumn}% Align table columns on decimal point
\usepackage{bm}% bold math

\newcommand{\ket}[1]{|#1\rangle}

\begin{document}

%\title{A polarization-independent phase shifter based on a Sagnac loop}% Force line breaks with \\
\title{Sagnac-loop phase shifter with polarization-independent operation}% Force line breaks with \\
\author{Tien Tjuen Ng, Darwin Gosal, Ant\'{\i}a Lamas-Linares and Christian Kurtsiefer\\
Centre for Quantum Technologies and Department of Physics,\\ National University of Singapore, 3 Science Drive 2, 117543 Singapore}% <-this % stops a space
%\thanks{The authors are with the Department
%of Physics and the Centre for Quantum Technologies, National University of Singapore, 117543 Singapore, e-mail: %christian.kurtsiefer@gmail.com.}% <-this % stops a space
%\thanks{Manuscript received April 19, 2005; revised January 11, 2007.}}
%\author{Tien Tjuen Ng}
%\author{Darwin Gosal}
%\author{Ant\'{\i}a Lamas-Linares}
%\author{Christian Kurtsiefer}
%\email{christian.kurtsiefer@gmail.com}
%\affiliation{Centre for Quantum Technologies / Department of Physics, 3 Science Drive 2, National University of Singapore, 117543 Singapore}

%\date{\today}
%\markboth{Journal of Quantum Electronics,~Vol.~X, No.~x, April~2008}%
{%Shell \MakeLowercase{\textit{et al.}}: Bare Demo of IEEEtran.cls for Journals}
\maketitle

%\begin{abstract}
%We present a bulk-optic, polarization-independent phase shifter which achieves
%a $\pi$ phase modulation in 1.6\,ns. A transverse mode electro-optic
%modulator using lithium niobate is placed in a Sagnac-like loop containing Faraday rotators. This direction dependent polarization-rotation in the loop is combined with a Mach-Zehnder interferometer to form an optical switch with an on/off contrast of 96\%. 
%\end{abstract}
\begin{abstract}
A bulk-optics, polarization-independent phase shifter for photonic quantum information applications is demonstrated. A transverse mode electro-optic modulator using lithium niobate is placed in a Sagnac-like loop containing Faraday rotators. This direction-dependent polarization rotation in the loop is combined with a Mach-Zehnder interferometer to form an optical switch with an on/off contrast of 96\% and a switching time of 1.6\,ns.
\end{abstract}

%\pacs{Valid PACS appear here} % PACS, the Physics and Astronomy
                             % Classification Scheme.


\IEEEpeerreviewmaketitle
\section{\label{sec:level1}Introduction}
Fast optical phase shifters or switches based on electro-optical materials are
in widespread use. Switch devices that operate on a sub-nanosecond timescale are a core building block of photonic technologies, in
particular in optical communications \cite{eldada:01,wooten:00}. For most
typical applications, the devices work for only one particular polarization
and can tolerate large insertion losses. The situation is very different in
quantum information science. Here, losses are much more important, and a
convenient degree of freedom to encode information is
the polarization of single photons, since it is particularly easy to manipulate and
measure. This implies that this degree of freedom should be left untouched by
any external operation, such as switching. The availability of a device which
can rapidly switch light with low losses and without affecting polarization
opens a new range of applications: conditional measurement schemes, combined
polarization state--time bin encoding, quantum computing with optical qubits
\cite{prevedel:07} and device--independent quantum cryptography \cite{acin:07}. For a qubit encoded in the polarization state of a photon,
the device would add an overall phase without affecting the quantum information
encoded. In Dirac notation, this is expressed by
\begin{equation}
\ket{\psi}=\alpha\ket{H}+\beta\ket{V} \to
e^{i\phi}[\alpha\ket{H}+\beta\ket{V}]=e^{i\phi}\ket{\psi}\,,
\end{equation}
where $\ket{H}$ and $\ket{V}$ represent horizontal and vertical polarization
states,
$\alpha$ and $\beta$ are the relative amplitudes, and $\phi$ is the global
phase. A device adding a global phase to a polarization qubit therefore must
preserve not only the relative amplitudes of the polarization components, but
also the relative phase. 

Electro-optic modulators (EOM) vary the phase of transmitted light with a
transverse or longitudinal electric field (with respect to the propagation
direction) by inducing a refractive index change in a crystal. While devices
based on a longitudinal field can be polarization independent, the typical
high driving voltages renders them
impractical for fast operation. To achieve an overall
non-birefringent operation of transverse EOMs, a series of
crystals can be combined \cite{davydov:07,burns:78,oh:00}. This makes fast
operation more difficult, and complicates the setups and
fabrication. Alternatively, the three electro-optic coefficients can be
simultaneously used \cite{kaplan:00}. Such elements, however, exhibit an
undesirable phase dependent beam deviation. Our aim was to make a device that
can be implemented in bulk optics, is appropriate for integration in typical
photonic quantum information setups, and can operate in a ns timescale
relevant to such experiments.

\begin{figure}
\includegraphics[width=\columnwidth]{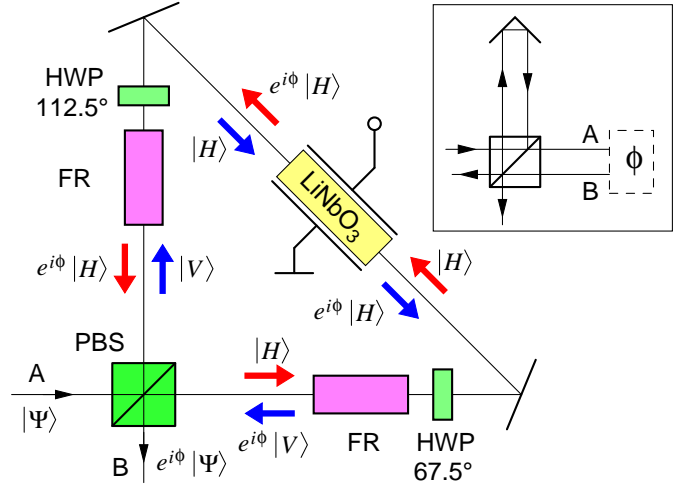}
\caption{\label{fig:setup}
%(Color online)
  Sagnac-like topology of a polarization-independent phase
  shifter. A polarizing beam splitter (PBS) divides an incoming light from
  port A into horizontal (H) and vertical (V) components. These
  counter-propagate through the loop, pass through a pair of Faraday rotators
  (FR) and half wave plates (HWP) surrounding a lithium niobate crystal. The two
  components thus acquire an identical phase $\phi$, are recombined by the PBS
  and exit through port B. The phase shifter is made part of a Mach-Zehnder
  interferometer (inset) to form an optical switch.}
\end{figure}

\section{Design Concept}
The basic idea  %behind our device
is to split the two polarization components
into two counterpropagating beams in a Sagnac-like loop (for a related device
based on a similar concept see~\cite{dennis:00}). These beams are then
manipulated such that in one section of the loop they have the same
polarization; in this section they both experience an identical phase shift
within a single modulator crystal. Further manipulation of the polarization
allows the recombination of the two counterpropagating beams such that the
original relative amplitudes of the polarization components are maintained and
a global phase has been added to the state.

The first step is to separate the two polarization components into
counterpropagating beams by a polarizing beam splitter (PBS). To ensure that
the two components have the same polarization at the modulator position in the
loop, we need to rotate the polarization of the clockwise beam by 90$^{\circ}$
twice, while the counterclockwise beam is not affected. This
direction-dependent rotation is achieved with a combination of Faraday
rotators (FR) and half wave plates (HWP). By situating the modulator crystal
in the symmetric point in the loop, we avoid geometric limitations in
operation speed from propagation speeds within the
device. Figure~\ref{fig:setup} schematically shows how the two
counterpropagating polarization components are split, manipulated, and
eventually recombined by the same PBS that performs the original splitting. 
In contrast to \cite{dennis:00}, the Faraday rotators in this design allow for
the phase shifted light to exit through a different port of the loop than the
input light.

\section{Implementation}
Lithium niobate is used as non-linear material for the modulator due to its
excellent optical and electro-optical properties
\cite{dmitriev:99, yariv:03, agullo-lopez:94}. To implement the basic idea with
free propagating beams we use an anti-reflection coated bulk 20\,mm long y-cut
LiNbO$_{3}$ crystal with a cross section of 1$\times$1\,mm$^2$ as a phase
shifting element. The y-cut is selected to avoid beam deviations induced by
the applied field.

The length of the crystal is chosen together with the cross section such that
the beam can propagate through the material within the Rayleigh length of a
transverse Gaussian mode, and the crystal is still reasonable easy to
fabricate and manipulate, while keeping a large aspect ratio, as
this results in a low half wave voltage.

%\section{\label{sec:level3}Implementation}
The polarization of light at the modulator was chosen to be parallel to the
optical axis for the lowest half wave voltage,
\begin{equation}
 V_{1/2}=\lambda\,d/\!L\,r_{33}\,n_e^3\,,
\end{equation}
where $d$ and $L$ are the thickness and length of the crystal, $\lambda$
the wavelength of the light, $n_{e}$ the extraordinary refractive index,
and $r_{33}$ the relevant electro-optic coefficient. From our crystal
geometry, we expect $V_{1/2}\approx100$\,V, which still can be switched with reasonable effort on a
time scale compatible with our photon counter jitter ($\approx1\,$ns).

\begin{figure}
\begin{center}
\includegraphics[width=\columnwidth]{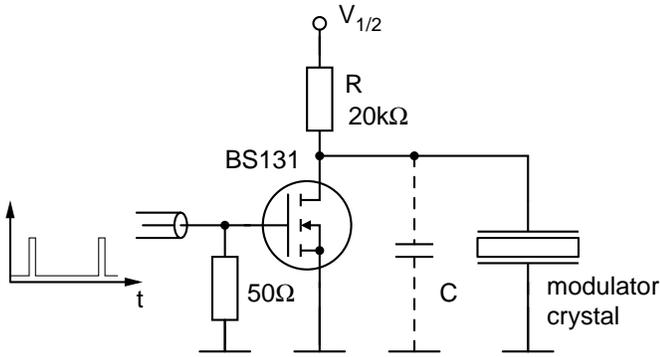}
\caption{Simple driving circuit to discharge the voltage across the Lithium
  niobate modulator in a time scale of nanoseconds.}
\label{circuit}
\end{center}
\end{figure}

For half wave voltages on this order, the driving circuit becomes an essential
part of the switch design (see Fig.~\ref{circuit}). The main component of the
driving circuit is a suitable MOSFET (BS131) located closely to 
the modulator crystal to keep parasitic capacitances small ($C=50$\,pF,
of which $\approx5\,$pF are from the crystal). While the MOSFET can
quickly discharge the crystal, a resistor $R=20\,$k$\Omega$ is 
used to recharge it again to the half wave voltage. Gate
pulses with a rise time of 400\,ps and an amplitude of 10\,V
were provided via an impedance-matched line.
While the switching time of the modulator for one direction is limited by the
discharge time of the crystal/MOSFET combination, the repetition rate
for switch changes has a much larger time constant $\tau=R\,C$ for
recharging the crystal. To maintain the full change of a half
wave voltage across the switch after a recharge, we limited the repetition rate
to 100\,kHz. 

This polarization-independent phase shifter is placed in a Mach-Zehnder
interferometer to test the performance of our device as a switch, as shown in
the small box in Fig.~\ref{fig:setup}. The interferometer is adjusted for zero
optical path length difference to allow for optimal operation over a large
range of wavelengths. The effects of wavefront mismatch of the two paths are
compensated via a divergent lens pair inserted in the reference path of the
interferometer. 

\section{\label{sec:level4}Experimental performance}
The performance of the switch was verified with a He-Ne laser at
632.8\,nm. Applying a low frequency saw tooth voltage to the crystal
electrodes, 
we checked the polarization-independency of the switch by measuring the
interferometer visibility $(I_{\rm on}-I_{\rm off})/(I_{\rm on}+I_{\rm off})$
as a measure for the on/off contrast $I_{\rm on}/I_{\rm off}$ for several
input polarization states. The 45$^\circ$ polarization is  
the most stringent test, since it clearly reveals any loss of a phase
relationship between the two polarization components. Table~\ref{tab:table1}
shows the 
measured the half wave voltage and the visibility for different input
polarizations of the laser beam after background correction. 

\begin{table}
%\caption{\label{tab:table1}Half wave voltage \& switch contrast for different
%  input polarization states.}
%\begin{ruledtabular}
\caption{\label{tab:table1} }
\centerline{
\begin{tabular}{c||c|c|c}
Polarization & $V_{1/2}\,{\rm (V)}$&Visibility&on/off contrast\\
\hline
$0^\circ$ & $93.9\pm 0.6$ & $96.1\pm1.7\%$&50.3 (17.0\,dB)\\
$45^\circ$& $94.6\pm 0.4$ & $93.3\pm1.6\%$&28.9 (14.6\,dB)\\
$90^\circ$& $94.9\pm 0.6$ & $94.7\pm1.7\%$&38.2 (15.8\,dB)
\end{tabular}
}
%\end{ruledtabular}
\end{table}

Currently, the overall visibility is limited by wavefront matching in the
interferometer. The reduced visibility of the 45$^\circ$ polarization is
caused by an imperfect polarization compensation of the beam splitting element
and the retro-reflectors in the interferometer.

The switching time of the phase shifter was characterized with a
fast photodiode (G5842, Hamamatsu) located at one of the interferometer
outputs; we observed a switching time (10\% to 90\% transition) of
$1.6\pm0.2\,$ns, following the limitation in the fall time of the
voltage across transistor and crystal (see Fig.~\ref{delay}).

The elements in the phase shifting loop (the two Faraday rotators and the LiNbO$_3$) are anti-reflection coated for operation in the near infrared. At the test wavelength of 633\,{\rm nm} we observe an insertion loss of $1\,{\rm dB}$, and we expect a loss of $0.2\,{\rm dB}$ at the optimal wavelength.

\begin{figure}
\begin{center}
\includegraphics[width=9cm,angle=0]{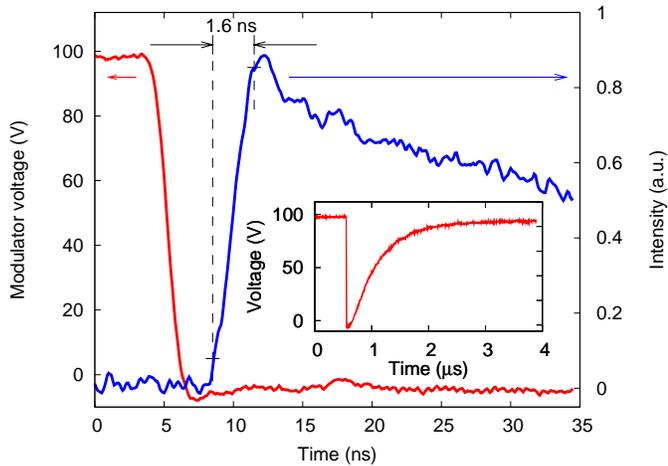}
\caption{
%(Color online) 
  Modulator voltage and light intensity at the output of the
  interferometer. The apparent delay between the voltage change and the
  response is a cable length artifact, and the decay of the light intensity 
  after switching is caused by the low frequency cutoff in the bias network of
  the high speed photodiode. The inset shows the recharging of the crystal,
  which is much slower,  and determines the maximum repetition rate with our
  driver.} 
\label{delay}
\end{center}
\end{figure}

\section{\label{sec:level5}Conclusions}
We reported on a simple polarization-independent electro-optic modulator with a
 switching time of 1.6\,ns which can be easily adapted to work with a range of different
wavelengths. The on/off contrast ratio was between 14.6 and 17\,dB, and showed
residual polarization effects, which can be attributed to imperfect
compensation of the external interferometer. Particular attention was paid to
testing the device in complementary polarization bases, as the preservation of
coherence between the polarization components is fundamental to our target
application. 

While currently implemented with free space optical components, the same basic
idea could be used with waveguides for applications where the larger insertion
loss is of less concern. There, switching times could be significantly
improved due to the greatly reduced half wave voltages. Such a device
has potential applications in photonic quantum communication experiments,
e.g. for carrying out conditional measurements on entangled multi-photon
states with polarization encoded qubits.

\section*{\label{sec:level6}Acknowledgments}
This work is supported by ASTAR SERC grant No. 052\,101\,0043, and the National
Research Foundation and Ministry of Education, Singapore.

%\begin{IEEEbiographynophoto}{Tien Tjuen Ng}
%Biography text here.
%\end{IEEEbiographynophoto}

\begin{thebibliography}{99}


\bibitem{eldada:01}Louay Eldada , ``Advances in telecom and datacom optical components'',
\emph{Opt. Eng.} , vol. 40, pp. 1165--1178, Jan. 2001.

\bibitem{wooten:00}Wooten E.L. et al.,``A review of lithium niobate modulators for fiber-optic communications systems'',
\emph{IEEE J. Sel. Top. Quantum Electron.}, vol. 6, pp. 69--82, Jan. 2000.

%\bibitem{10}M. A. Nielsen and I. L. Chuang,
%\emph{Quantum Computation and Quantum Information} {(Cambridge University Press)}, (2000)

\bibitem{prevedel:07}Prevedel et al.,``High-speed linear optics quantum computing using active feed-forward'',
\emph{Nature}, vol. 445, pp. 65--69, Jan. 2007.

\bibitem{acin:07} A. Acin et al., ``Device-independent security of quantum cryptography against collective attacks'', \emph{Phys. Rev. Lett.} vol. 98, 230501, 2007.  

\bibitem{davydov:07}B. L. Davydov, A. A. Krylov, D. I. Yagodkin, ``Polarisation-independent electrooptical switch based on LiNbO$_3$ and LiTaO$_3$ crystals'', \emph{Quantum Electron.} vol. 37, pp. 484--488, 2007.

\bibitem{burns:78}W. K. Burns, T. G. Giallorenzi, R. P. Moeller, and E. J. West, ``Interferometric waveguide modulator with polarization-independent operation'', \emph{Appl. Phys. Lett.}, vol. 33, pp. 944--947, Dec. 1978.

\bibitem{oh:00}H.-H. Oh, S.-W. Ahn, and S.-Y. Shinn, ``Polarisation-independent phase modulator using electro-optic polymer'',
\emph{Electron. Lett.} vol. 36, pp. 969--970, May 2000.

\bibitem{kaplan:00}Arkady Kaplan, and Shlomo Ruschin, ``Layout for polarization insensitive modulation in LiNbO$_3$ waveguides'',
\emph{IEEE J. Sel. Top. Quantum Electron.}, vol. 6, pp. 83--87, Jan. 2000.

\bibitem{dennis:00}M. L. Dennis and I. N. Duling, III, ``Polarisation-independent intensity modulator based on lithium niobate'', \emph{Electron. Lett.}, vol. 36, pp. 1857--1858, Oct. 2000.

%\bibitem{wooten:00}E. L. Wooten et al.,
%\emph{IEEE J. Sel. Top. Quantum Electron.} {\bf 6}, 69 (2000), and references therein

\bibitem{thylen:88}L. Thylen, ``Integrated optics in LiNbO$_3$: recent developments in devices for telecommunications'',
\emph{J. Lightwave Technol.}, vol. 6, pp. 847--861, Jun. 1988.

\bibitem{dmitriev:99}V. G. Dmitriev, G. G. Gurzadyan, and D. N. Nikogosyan,
\emph{Handbook of Nonlinear Optical Crystals} {\bf 64}, 119 (1999).

\bibitem{yariv:03}A. Yariv, P. Yeh,
\emph{Optical Waves in Crystals: Propagation and Control of Laser Radiation} {(John Wiley \& Sons)}, (2003).

\bibitem{agullo-lopez:94}Fernando Agullo-Lopez et al.,
\emph{Electrooptics: Phenomena, Materials And Applications} {(Academic Press)}, 138 (1994).

\end{thebibliography}
\end{document}